\documentclass[prl,aps,amsfonts,amssymb,amsmath,graphicx,showpacs,twocolumn,epsfig,amscd,ulem]{revtex4}

\usepackage{graphicx}
\usepackage{natbib}
\usepackage{amssymb}
\usepackage{amsmath}
\usepackage{verbatim}

\begin{document}

\title{Fluctuation-dissipation theorem in an aging colloidal glass}

\author{Sara Jabbari-Farouji $^{1}$, Daisuke Mizuno $^{2}$,
Maryam Atakhorrami$^{2}$, Fred C.\ MacKintosh$^{2}$, Christoph F.\
Schmidt$^{2}$, Erika Eiser$^{3}$, Gerard H.\ Wegdam $^{1}$ and
Daniel Bonn$^{1,4}$}

\affiliation{$^{1}$Van der Waals-Zeeman Institute, University of
Amsterdam, 1018XE Amsterdam, the Netherlands }

\affiliation{$^{2}$ Department of Physics and Astronomy, Vrije
Universiteit Amsterdam, 1081HV Amsterdam, The Netherlands }

\affiliation{$^{3}$van't Hoff Institute for Molecular Sciences,
University of Amsterdam, 1018WV Amsterdam, The Netherlands }

\affiliation{$^{4}$Laboratoire de Physique Statistique, UMR 8550
CNRS associated with the University of Paris 6 and Paris 7, Ecole
Normale Sup\'erieure, 75231 Paris Cedex 05, France  }

\date{\today}

\begin{abstract}
We provide a direct experimental test of the
fluctuation-dissipation theorem (FDT) in an aging colloidal glass.
The use of combined active and passive microrheology allows us to
independently measure both the correlation and response functions
in this non-equilibrium situation. Contrary to previous reports,
we find no deviations from the FDT over several decades in
frequency (1 Hz-10 kHz) and for all aging times. In addition, we
find two distinct viscoelastic contributions in the aging glass,
including a nearly elastic response at low frequencies that grows
during aging.
\end{abstract}

\maketitle

Developing a statistical mechanical description of non-equilibrium
systems such as glasses still remains an important challenge in
physics. One of the most interesting recent developments along
these lines is the proposal to generalize the fluctuation
dissipation theorem (FDT) to non-equilibrium situations
\cite{Cuglian1}. The FDT relates the response of a system to a
weak external perturbation to the relaxation of the spontaneous
fluctuations in equilibrium \cite{Landau}. The response function
is proportional to the power spectral density of thermal
fluctuations, with a prefactor given by the temperature. This
suggests a generalization for systems out of equilibrium, in which
the (non-equilibrium) fluctuations are related to the response via
a time-scale-dependent \emph{effective temperature}. While this
has been studied extensively theoretically \cite{simulation}, the
experimental support for a meaningful effective temperature is
unclear. There have been few experiments
\cite{Israel,Ciliberto,abou} and the usefulness of the extension
of the FDT to non-equilibrium situations is still a matter of
controversy.

Here, we introduce a combination of both \emph{active} and
\emph{passive} (fluctuation-based) microrheology techniques
\cite{Weitz,Gittes2,active} that provide a way to \emph{directly}
test the applicability of the FDT. We examine the validity of FDT
in a colloidal glass, the synthetic clay of Laponite
\cite{Bonn2,Bellour,glass}. For this system conflicting results
have been reported previously \cite{Ciliberto,abou}, that may in
part be due to the use of a limited experimental window in both
frequency and aging time. Here, we perform measurements over a
wide range of frequencies and aging times. Contrary to previous
reports, we find no violation of the FDT and thus no support for
an effective temperature different from the bath temperature.

In addition, these measurements provide a new insight into the
physics of the aging process. The microrheology done during the
aging process allows us to explore the evolution of viscoelastic
properties of the glass over a wider frequency range than hitherto
explored, spanning nearly 6 decades in frequency. The measurements
reveal the existence of two distinct contributions to the
viscoelasticity of the system: (i) a high-frequency viscoelastic
response in which the shear modulus increases rapidly with
frequency; and (ii) a predominantly elastic (weakly
frequency-dependent) response at lower frequencies, which becomes
increasingly important as the system ages.

The Einstein relation relates the diffusion of the particle
(\emph{i.e.}, position fluctuations) to its mobility. This is a
special case of the FDT, and its generalization to a viscoelastic
system in Fourier space relates the power spectral density (PSD)
of position fluctuations to the imaginary part of the complex
response function $\alpha''(\omega)$ \cite{Weitz,Gittes2}:
\begin{equation}
\label{eq:a1}  \langle|x(\omega)|^{2}\rangle=
\frac{2k_{B}T}{\omega}\alpha''(\omega)
\end{equation}
Here, $\langle|x(\omega)|^{2}\rangle$ denotes the Fourier
transform of the ensemble average $\langle x(t)x(0)\rangle$. In a
non-equilibrium system this suggests the introduction of an
effective temperature in which $T$  is replaced by
$T_{\mbox{\scriptsize eff}}(\omega)$ in Eq.\ (\ref{eq:a1}).

To investigate the aging of colloidal glass, we study the motion
of probe particles using optical tweezers. A custom-built inverted
microscope \cite{Maryam}, equipped with two overlapping optical
tweezers formed by two independent lasers (wavelengths $ \lambda$=
830 nm and 1064 nm ) focused to diffraction-limited spots. The
latter drives the oscillations of the trapped particle with an
Acousto-Optical Deflector, allowing us to measure the (active)
response to a driving force. The ($x$, $y$) position of the
particle is determined by a quadrant photo diode \cite{Gittes3}
with a spatial resolution of $\sim 0.1$ nm. The output signal from
the photodiode is fed into a lock-in amplifier that measures the
amplitude and phase of the particle displacement $x(t)$ caused by
an oscillatory motion of the drive laser focus. From the motion of
the drive laser we determine the force $F(t)$ acting on the
particle. The response function is then given by $\alpha(\omega) =
x_{\omega}/F_\omega$, where $x_{\omega}$ and $F_\omega$ denote the
fourier transforms of $x(t)$ and $F(t)$. By measuring the PSD of
the same beads under the same conditions in water, we are able to
calibrate both trap stiffness \cite{Gittes2,Maryam} and particle
displacement for active and passive data \cite{Gittes2}.

For the passive measurements the shutter in front of the driving
laser is closed and the spontaneous fluctuations of the particle
position are recorded for a minimum time of 45 s. From the
displacement time series, we calculate the displacement power
spectral density by Fast Fourier Transform \cite{Gittes2}.
Comparing the response function from the active microrheology with
the fluctuation spectra, we can directly check the validity of the
FDT during the aging of the system, as well as resolve the
frequency-dependent viscoelastic properties during the aging of
the glass.

\begin{figure}
\includegraphics[scale=0.75]{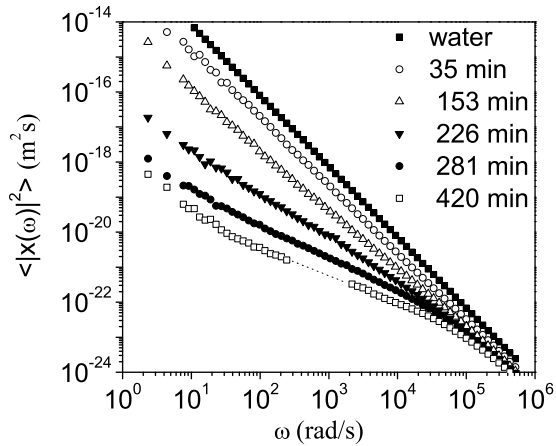}
\caption{ \label{fig1}The displacement power spectral densities
(PSD) as a function for frequency of 1.16 $\mu$m silica probe
particles with increasing age after preparing the sample.
Fluctuations were recorded for 45 seconds with the 830 nm laser
focus and results averaged  in x-y directions for 1 bead several
times. Aging times are given in the legend. The filled squares
show the PSD of a bead in pure water for comparison. An acoustic
noise signal around $f\approx$ 200Hz is cut out from the curve at
the latest stage of aging where the displacement signal was the
lowest. All experiments were done at $21^{o}C$.}
\end{figure}

The colloidal glass under study is a suspension of Laponite XLG in
ultra pure water. After mixing the colloidal particles with the
water, the system spontaneously evolves from an initially liquid
and ergodic state to a non-ergodic glassy state that exhibits
elastic behavior \cite{Bonn2}. For a particle concentration of 2.8
wt\%, the rate of aging is slow enough, on the one hand, that no
significant structural and dynamic changes occur during each
individual active and passive microrheology measurement lasting at
most 2 min. Nevertheless, the system evolves fast enough to allow
us to follow the evolution from `liquid' to `solid' (the glass no
longer flows  when the sample cell is tilted) within about 8
hours. The dispersions are filtered (Millipore Millex AA $0.8$
$\mu$m filter) to obtain a reproducible initial state
\cite{glass}. This defines the zero of aging time $t_{a}=0$.
Immediately after filtration, a small fraction ($<10^{-4}$ vol\%)
of silica probe beads (diameter 1.16 $\mu$m $\pm 5 \%$) are mixed
with the Laponite dispersion. The solution is then introduced into
a sample chamber of about 50 $\mu$l volume, consisting of a
coverslip and a microscope slide separated by a spacer of
thickness 70 $\mu$m. This is sealed with vacuum grease to avoid
evaporation of sample. We then trap a single silica bead and
perform the active and passive experiments on it.

Since the system evolves toward a non-ergodic state, the time
average may not be equal to ensemble average for the measured
PSDs. To investigate this, we confirmed that reproducible PSDs
were obtained for the same aging time, independent of bead
position, during all stages of aging. We also confirmed that our
results do not depend on the time interval used to compute the
time average. Thus, we can use the time-averaged PSD without
averaging over several beads in our study. Figure \ref{fig1} shows
the (passive) displacement PSD for different aging times. It is
evident that the particle motion progressively slows down with
increasing aging time, reflecting the increase in viscosity of the
system. Qualitatively two regimes of aging are seen: for $t_{a}<
200$ min the PSD can be described by a single power law. At longer
aging times two distinct slopes appear in the log-log plot (Fig.\
\ref{fig1}).

\begin{figure}
\includegraphics[scale=0.85]{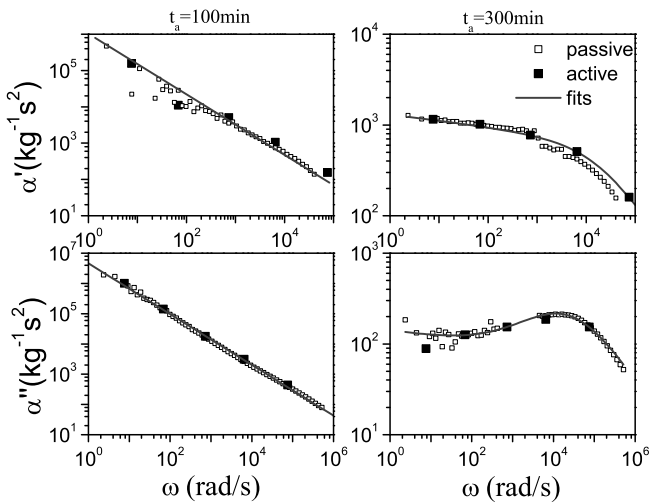}
\caption{ \label{fig2} Comparison of active and passive results:
Real $\alpha'(\omega)$ and imaginary $\alpha''(\omega)$ at
$t_{a}=100 $ and 300 min obtained from active (filled symbols) and
passive (open symbols) microrheology performed on the same 1.16
$\mu$m diameter silica bead in the same sample. For the passive
experiments, the imaginary parts of the response functions are
obtained directly and real parts are calculated with a
Kramers-Kronig integral. The lines show the fits to
Eq.~(\ref{eq:a2}). At early stages of aging the data can be
described with one power law, while at later stages, a
superposition of two power laws is needed to describe the whole
frequency range. The amplitude of oscillation for the active
experiments was 77 nm.}
\end{figure}

We also measure the (active) response of the same bead used in
passive measurements, as a function of aging time and for
oscillation frequencies of $f$ =1.2, 10.8, 116, 1035 and 12000 Hz.
To directly compare the (passive) fluctuations with the (active)
response, we express our fluctuation PSDs normalized in such a way
as to permit a direct comparison with the measured $\alpha$ in the
form of Eq.\ (\ref{eq:a1}). Thus, we plot the measured PSD
multiplied by $\omega/(2k_BT)$. We obtain the real part using a
Kramers-Kroning (principal-value) integral \cite{Landau}
$\alpha'(\omega)=\frac{2}{\pi}P\int^\infty_0 \frac{\xi \alpha
''(\xi)}{\xi^{2}-\omega^{2}} d \xi$. The cutoff error due to a
finite range of frequencies sets an upper limit to
$\alpha'(\omega)$ about a decade lower than that of
$\alpha''(\omega)$. Fig.~\ref{fig2} depicts the real and imaginary
parts obtained from the active and passive methods at an early
$(t_{a}=100 $ min) and a late stage of aging $(t_{a}=300 $ min).
We see that the results are identical to within the experimental
accuracy, showing that there are no deviations from the FDT in
this system over the range of frequencies and aging times probed
in our experiments. Note that the small deviations between the
respective $\alpha'(\omega)$ values at high frequencies are likely
due to cut-off errors in the Kramers-Kronig integrals because the
drawn lines, obtained by extrapolating the measured power law of
$\alpha ''$ to infinity, show significantly better agreement.
Since we have directly compared both the real and imaginary parts
of the response functions, this represents a stronger test of the
FDT than previous measurements \cite{abou} and demonstrates that
the FDT holds in this non-equilibrium system.

\begin{figure}
\includegraphics[scale=1]{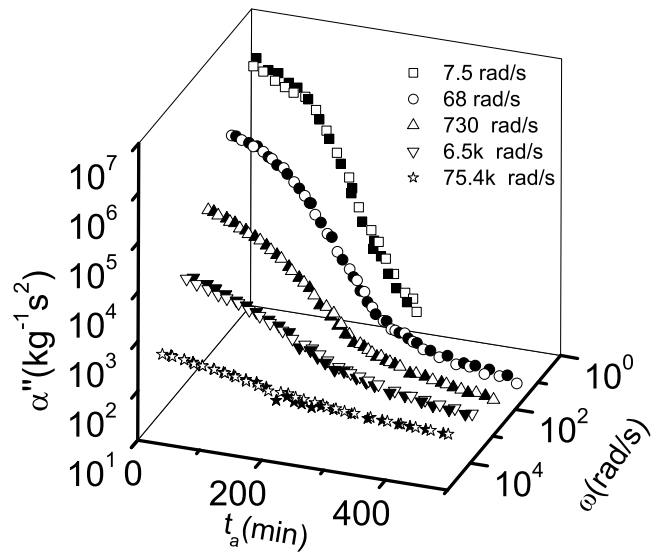}
\caption{The comparison of $\alpha''$ extracted from passive (open
symbols) and active (filled symbols) measurements as a function of
aging time. For the lowest frequency, we could not measure
reliable data longer than 300 min, since the signal to noise to
ratio became of the order of 1 with increasing aging time, as the
material became stiff.}
 \label{fig3}
\end{figure}

In Fig.~\ref{fig3}, we plot the extracted $\alpha''$ as a function
of aging time for several different frequencies. As can be seen
the active and passive data agree very well. This figure confirms
again that to within the experimental uncertainty the FDT holds:
the measured effective temperature does not differ from the bath
temperature.  The resulting effective temperature
${T_{\mbox{\scriptsize eff}}}/{T_{\mbox{\scriptsize
bath}}}={\alpha_{\mbox{\scriptsize
passive}}''}/{\alpha_{\mbox{\scriptsize active}}''}$ is shown in
the Table I. We conclude that the measurements show the FDT in the
form of the Einstein relation is valid for all frequencies probed
here, and can be used for all the stages of aging in this system.
\begin{table}\label{table1}
\centering
\begin{tabular*}{8.5cm}[b]{|l|l|l|l|l|l|}\hline
\multicolumn {6}{|c|}{\bfseries ${T_{\mbox{\scriptsize
eff}}}/{T_{\mbox{\scriptsize bath}}}$}\\ \hline
$t_{a}$ & 7.5 rad/s & 68 rad/s & 728 rad/s & 6.5 krad/s & 75 krad/s \\
\hline
 0-2 h &  0.75 $\pm$ 0.3 & 1$\pm$ 0.1 & 0.95 $\pm$ 0.1 & 0.85$\pm$ 0.1 & 1.0 $\pm$ 0.1\\
 2-4 h &  1.2 $\pm$ 0.3 & 1 $\pm$0.1 & 1 $\pm$ 0.1 & 0.9 $\pm$ 0.1 & 1.0 $\pm$ 0.1\\
4-6 h &  1.4 $\pm$ 0.3 & 1 $\pm$0.1 & 1.1 $\pm$ 0.1 & 1.1$\pm$
0.1 & 1.1 $\pm$ 0.1\\
 6-8 h &  & 0.85 $\pm$0.1 & 1.0 $\pm$ 0.1 & 1.1 $\pm$ 0.1 & 1.0 $\pm$
 0.1\\ \hline
\end{tabular*}\caption{The effective temperature obtained for different
frequencies averaged over 2 h time intervals. Within the
uncertainty in the experiments, ${T_{\mbox{\scriptsize
eff}}}/{T_{\mbox{\scriptsize bath}}}=1$  }
\end{table}
The method also allows us to obtain the viscoelastic properties
over a very wide frequency range; classical (macroscopic) rheology
is limited to frequencies up to about 100 rad/s \cite{Bonn2}. In
Fig.\ \ref{fig1} we observe a gradual decrease of the PSD for
higher frequencies and a more rapid change at lower frequencies.
The response function is directly proportional to the PSD which in
turn should be inversely related to the complex shear modulus
$G^{*}=G'+ \imath G''$. With increasing aging time, as the sample
becomes more viscous and solid-like, one would expect that both
the elastic modulus $G'$ and the viscous modulus $G''$ increase.
This is consistent with the decrease of the PSD and consequently
of the response function. As was mentioned above, at late stages
of aging two distinct slopes appear in the PSD (Fig.\ \ref{fig1}).
This suggests the existence of two distinct contributions to the
viscoelasticity during aging. Assuming the generalized Stokes
formula for the viscoelastic response function
\cite{Weitz,Gittes2}, we obtain excellent fits to the data
assuming a simple addition of two power law contributions to the
complex shear modulus (only a single power-law contribution at the
early stages of aging):
\begin{equation}
\label{eq:a2}\alpha(\omega)=\frac{1}{6 \pi R
G^{*}(\omega)}=\frac{1}{C_{1}(-\imath \omega)^{a}+C_{2}(-\imath
\omega)^{b}}
\end{equation}

\begin{figure}
\includegraphics[scale=0.80]{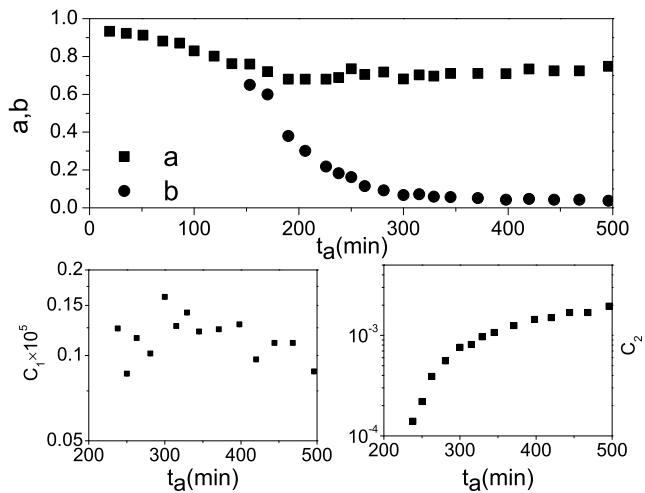}
\caption{ \label{fig4}For $t_{a}< $ 150 min, the complex response
function can be described by single power law of the form
${1}/{(-\imath \omega)^{a}}$ . After about 150 min, a slowly
relaxing contribution emerges leading to a response function of
the from in Eq.\ (2).}
\end{figure}

The fit of the imaginary part of the response function from the
passive measurements with the imaginary part of the above
functional form is shown by the drawn lines in Fig.\ \ref{fig2}.
To demonstrate the quality of this model for describing the data,
we also plot the real part with the fitting parameters obtained
from the imaginary part. The agreement is remarkable, especially
with the active data at high frequencies. Fig.\ \ref{fig4} depicts
the evolution of the fitting parameters, \emph{i.e.}, the
exponents of power laws and the weight factors for the
contribution of the two viscoelastic contributions as a function
of aging time. The exponent and amplitude of one of the components
do not change with aging time while the amplitude of the other one
grows appreciably for aging times longer than about 250 min.

These results demonstrate the existence of two distinct
contributions in the viscoelasticity of the system. In addition to
a strongly frequency-dependent viscoelastic response at high
frequencies, we also observe the slow development of a more
elastic (weakly frequency-dependent) response during the aging. In
fact, this appears to be the main characteristic of the aging in
this system. A similar description in terms of a network in a more
fluid-like background has been suggested before for polymeric gels
\cite{twofluid,Trappe}. In our case, this can be attributed to the
growth of a tenuous network-like structure, in addition to the
more viscous response that is always present. We note, however,
that although our system also becomes non-ergodic there is no
evidence for spatial inhomogeneity, neither in our system, nor in
glassy systems in general. Recently, it has been suggested that
glassy systems may be dynamically heterogeneous
\cite{heterogeniety}, and that one should look into the
correlation of spatial and temporal dynamics in order to detect
the heterogeneity. Therefore the homogeneity of the combined
spatial and temporal dynamics merit further detailed study.

In summary, we see a good quantitative agreement between the
response function and the spontaneous thermal fluctuations,
implying that we observe no violation of the FDT in this
non-equilibrium system. Equivalently, we find an effective
temperature that does not differ from the system temperature. It
is important to note that these measurements provide a much more
direct test of the FDT than prior experiments, since we directly
measure the response and the \emph{corresponding} fluctuations
over the same and very large range of frequencies.

\textbf{Acknowledgments}

The research has been supported by the Foundation for Fundamental
Research on Matter (FOM), which is financially supported by
Netherlands Organization for Scientific Research (NWO). LPS de
l'ENS is UMR8550 of the CNRS, associated with the universities
Paris 6 and 7. We would like to thank N. Israeloff and H. Tanaka
for helpful discussions.

\end {document}